\pdfoutput=1

\documentclass[11pt]{article}
    
\usepackage[preprint]{coling}

\usepackage{arabtex}
\usepackage{utf8}
\setcode{utf8}
\usepackage{hyperref}
\usepackage{multirow}
\usepackage{enumitem}
\usepackage{listings}
\usepackage{amsmath}
\usepackage{algorithmic}
\usepackage{graphicx}
\usepackage{textcomp}
\usepackage{xspace}

\usepackage{times}
\usepackage{latexsym}
\usepackage[T1]{fontenc}
\usepackage[utf8]{inputenc}

\usepackage{microtype}
\usepackage{inconsolata}
\definecolor{blue}{HTML}{2f4b7c}
\definecolor{dkgreen}{rgb}{0,0.6,0}
\definecolor{gray}{rgb}{0.5,0.5,0.5}
\definecolor{mauve}{rgb}{0.58,0,0.82}
\definecolor{lightred}{RGB}{237, 67, 55}
\definecolor{lightviolet}{RGB}{215, 75, 118}

\definecolor{codegreen}{rgb}{0,0.6,0}
\definecolor{codegray}{rgb}{0.5,0.5,0.5}
\definecolor{codepurple}{rgb}{0.58,0,0.82}
\definecolor{backcolour}{rgb}{0.95,0.95,0.92}
\lstdefinestyle{mystyle}{
    backgroundcolor=\color{backcolour},
    commentstyle=\color{codegreen},
    keywordstyle=\color{blue},
    numberstyle=\tiny\color{codegray},
    stringstyle=\color{codepurple},
    basicstyle=\footnotesize\ttfamily,
    breakatwhitespace=false,
    breaklines=true,
    captionpos=b,
    keepspaces=true,
    numbers=left,
    numbersep=5pt,
    showspaces=false,
    showstringspaces=false,
    showtabs=false,
    tabsize=2
}

\newcommand{\dataname}{\emph{Dialect2SQL}\xspace}

\AtBeginDocument{%
  }

\title{\dataname  : A Novel Text-to-SQL Dataset for Arabic Dialects with a Focus on Moroccan Darija}

\author{
 \textbf{Salmane Chafik\textsuperscript{1}},
 \textbf{Saad Ezzini\textsuperscript{2}},
 \textbf{Ismail Berrada\textsuperscript{1}},
 \\
 \textsuperscript{1}Mohammed VI Polytechnic University, Ben Guerir, Morocco \\
 \textsuperscript{2}King Fahd University of Petroleum and Minerals, Saudi Arabia \\
 \small{
 chafik.salmane@um6p.ma, saad.ezzini@kfupm.edu.sa,
 ismail.berrada@um6p.ma
 }
}
\begin{document}
\maketitle

{\centering\large\textbf{Abstract}\par}
\vspace{0.5em}
\begingroup
\setlength{\parindent}{0.6cm}
\setlength{\rightskip}{0.6cm}
\setlength{\leftskip}{0.6cm}
\noindent The task of converting natural language questions (NLQs) into executable SQL queries, known as text-to-SQL, has gained significant interest in recent years, as it enables non-technical users to interact with relational databases. Many benchmarks, such as SPIDER and WikiSQL, have contributed to the development of new models and the evaluation of their performance. In addition, other datasets, like SEDE and BIRD, have introduced more challenges and complexities to better map real-world scenarios. However, these datasets primarily focus on high-resource languages such as English and Chinese. In this work, we introduce \dataname, the first large-scale, cross-domain text-to-SQL dataset in an Arabic dialect. It consists of 9,428 NLQ-SQL pairs across 69 databases in various domains. Along with SQL-related challenges such as long schemas, dirty values, and complex queries, our dataset also incorporates the complexities of the Moroccan dialect, which is known for its diverse source languages, numerous borrowed words, and unique expressions. This demonstrates that our dataset will be a valuable contribution to both the text-to-SQL community and the development of resources for low-resource languages.
\par
\endgroup

\texttt{\textbf{Keywords :}} Text-to-SQL, Low Resource Language, Moroccan Dialect

\section{Introduction}
SQL or Structured Query Language is a powerful, standardized programming language used by developers to interact with relational databases. It provides a framework for defining, manipulating, and querying data stored in a structured format, typically organized into tables. It is essential for managing the creation, retrieval, update, and deletion of data, commonly referred to as CRUD operations (Create, Read, Update, Delete). SQL is commonly used in various applications, from small systems to large-scale enterprise platforms, and is integral to desktop, web, and mobile applications alike. Mastery of SQL remains a foundational skill for software engineers and professionals working with databases and data management.

Implementing SQL queries has become significantly easier and simpler with the introduction of text-to-SQL models, which can convert natural language questions (NLQs) into executable and efficient SQL queries \cite{survey_2022}. The availability of various datasets and benchmarks, such as \cite{spider_2019, wikisql_2017}, has facilitated the training, fine-tuning, and evaluation of code-based Large Language Models (LLMs) for the text-to-SQL task.

The development of such datasets and models was driven by the significant demand for text-to-SQL chatbots and integrated applications, which provide an environment for generating and executing SQL queries. These tools allow non-technical users, who may not be familiar with SQL, to interact with a deployed relational database using everyday language. Such applications have immense potential across industries that store data in structured formats and make it accessible to users via web or mobile applications. For example, in the healthcare sector, text-to-SQL integrated applications can enable doctors and other medical professionals to easily query patient records or retrieve statistics by simply asking questions like, 'How many patients had advanced-stage cancer in 2025 and survived?', all without needing SQL knowledge. This capability not only saves time but also provides crucial data insights that can inform patient care and treatment planning. Similarly, in the finance sector, a financial analyst could ask, 'What was the revenue growth for each quarter this year?' and retrieve relevant data directly from a financial database. This simplifies data analysis and allows analysts to focus on interpretation rather than query composition.

However, previous work has primarily focused on high-resource languages, such as English and Chinese, often by translating English versions of these datasets. While translation models have significantly improved for high-resource languages, creating text-to-SQL datasets for low-resource languages or dialects remains challenging. This difficulty stems from the need for skilled software engineers who not only fully understand SQL syntax but also have a strong command of English, as most existing resources and dataset examples are in English. Additionally, cultural and linguistic differences can affect how questions are phrased, making it difficult to adapt high-resource or even multilingual text-to-SQL models to these languages and dialects.

To address these challenges, we introduce what we believe to be the first text-to-SQL dataset specifically developed for an Arabic dialect, named \dataname. This dataset is tailored to the Moroccan dialect, also known as Darija, which is known by its linguistic complexity. Moroccan Darija is a unique mix, incorporating vocabulary and grammatical structures from a diverse range of source languages, including Arabic, Berber, French, and Spanish. It features numerous borrowed words and distinctive expressions that set it apart from Modern Standard Arabic and other Arabic dialects, making it particularly challenging for natural language processing tasks. We believe that \dataname will play a significant role in advancing text-to-SQL capabilities for low-resource languages.

The paper is structured as follows. Section \ref{Sec:related-work} presents a review of related work, while Section \ref{Sec:Approach} provides a detailed explanation of each step involved in the construction of \dataname. We finish concluding the paper and suggesting potential directions for future research.

\section{Related Work} \label{Sec:related-work}

In recent years, there has been significant progress in the field of text-to-SQL. Various studies \cite{survey_2022, survey_2023} focused on improving the accuracy and efficiency of converting natural language questions into SQL queries, and others focused on addressing the critical needs of datasets and benchmarks.

Zhong et al. \cite{wikisql_2017} introduced the first large-scale cross-domain text-to-SQL dataset WikiSQL, composed of 80,654 examples distributed across 24,241 tables from Wikipedia in different domains. However this dataset was judged of simplicity, each question concerns only one simple table. To address this problem, 11 students from Yale University manually annotated a text-to-SQL dataset named SPIDER  \cite{spider_2019}. This dataset comes with more complex queries joining multiple tables and spanning different domains and databases. However, both datasets were judged non-realistic because of the way they were created, simple database schemas, and simple questions.

To address this issue, hazoom et al. \cite{sede_2021} introduced SEDE, a text-to-SQL dataset dedicated solely for training and evaluation, composed of 12,023 NLQ-SQL pairs collected from real usage on the Stack Exchange website, including a variety of real-world challenges rarely reflected in previous works. In the same context, Li et al. \cite{bird_2023} constructed another benchmark named BIRD containing 12,751 pairs, 95 databases, and spanning over 37 professional domains. This benchmark comes with more challenges to immitate real-world situation by providing long sequence schemas, One database may include up to 60 tables, and dirty values.

While these studies focused on English datasets, other works have explored datasets in additional languages. For example, Dou et al. \cite{multi_spider_2022} manually translated the SPIDER dataset into multiple languages, including English, German, French, Spanish, Japanese, Chinese, and Vietnamese. They conducted various experiments using multilingual models in each language to assess the impact of training large language models on the same dataset across different languages simultaneously. Additionally, they introduced a framework called SAVE (Schema Augmentation with Verification) to help close the performance gap between models trained on the English dataset and those trained on other languages.

On the other hand, Bakshandaeva et al. \cite{pauq_2022} introduced PAUQ, the first Russian text-to-SQL dataset, which they developed based on the SPIDER dataset. They trained two baseline models, RAT-SQL \cite{wang2019rat} and BRIDGE \cite{lin2020bridging}, on PAUQ to assess the trade-offs between using automatically translated and manually crafted natural language questions. Their analysis highlights the strengths and limitations of each approach, offering insights into how translation quality affects model performance in multilingual text-to-SQL tasks.

Similarly, Almohaimeed et al. \cite{ar_spider_2024} introduced an Arabic version of the SPIDER dataset, naming it Ar-SPIDER. To explore the linguistic challenges specific to Arabic, the authors fine-tuned two base models, LGESQL \cite{cao2021lgesql} and S2SQL \cite{hui2022s}, using two different multilingual encoders: mBERT \cite{pires2019multilingual} and XLM-R \cite{conneau2019unsupervised}. Additionally, they proposed a Context Similarity Relationship (CSR) approach, which led to a significant increase in overall performance, helping to close the gap between Arabic and English language models.

Other datasets have been created from scratch to support cross-database context-dependent Text-to-SQL (XDTS) tasks. For instance, the CHASE dataset \cite{chase_2021} includes 17,940 questions in Chinese designed specifically for XDTS. CHASE enables models to handle complex, multi-turn questions across different databases, facilitating research into both cross-database adaptability and contextual dependency in query generation. Likewise, the SeSQL dataset \cite{sesql_2022} comprises 27,012 question-SQL pairs, also in Chinese. SeSQL further enriches the resources available for training and evaluating models on XDTS tasks by providing a wide array of question types and database contexts.

Motivated by these works, our paper introduces a large-scale, cross-domain text-to-SQL dataset in the Moroccan dialect, based on the well-known BIRD dataset \cite{bird_2023}.

\section{Approach} \label{Sec:Approach}

This section explains the choice of dataset, the translation process, and presents key statistics for \dataname.

\subsection{Dataset}

The BIRD dataset, formally known as the BIg Bench for laRge-scale Database Grounded Text-to-SQL Evaluation \cite{bird_2023}, represents one of the latest and most comprehensive resources for evaluating text-to-SQL systems. Released at the end of 2023, BIRD is designed to test the capabilities of models in generating SQL queries from natural language questions across a diverse set of domains and databases. It contains 12,751 unique question-SQL pairs, which span across 95 extensive databases in 37 distinct domains.

We chose BIRD because of the unique challenges it introduces. This dataset includes long schemas, with some databases containing up to 60 tables. It also incorporates dirty values, where natural language questions may include incomplete or abbreviated values. In such cases, the model must infer the correct values using external knowledge, a new aspect introduced by this dataset. Additionally, BIRD features complex queries that may join up to six tables in a single query and utilize various functions not seen in previous datasets.

\subsection{Dataset Translation}

To achieve an efficient translation, we use GPT-4 to translate BIRD questions of the train set into Moroccan Darija. We then ask three computer science students, one PhD student and two master’s students, who are native speakers of Moroccan Darija and proficient in SQL, to edit these questions according to the following guidelines:

\begin{itemize} 
    \item The English question is translated into Darija using Arabic letters. 
    \item Values such as names, surnames, countries, cities, company names, and movie titles remain in English. 
    \item Numbers are written using the Hindu-Arabic numeral system, or Western Arabic numerals (1, 2, 3) rather than Eastern Arabic numerals (\<١, ٢, ٣>). 
    \item The context for this SQL task, which includes table-creation statements (e.g., \texttt{CREATE TABLE ...}), is not translated. 
\end{itemize}

The first guideline was established because many Moroccans use Latin characters to write in Darija. To avoid confusion, we implemented this guideline. The second guideline was created because personal or company names can be written in various ways using Arabic letters. For example, the name "Wolfgang Reitherman" can be written in different forms, as shown in Table \ref{tab:variants}. The back translation to English might change a letter or two, which can lead to different results in an SQL query. The final guideline was established because the context is an SQL query that creates database tables including columns and their types, that's why it should remain in SQL (English). 

A final iteration was conducted by the same PhD student to ensure the quality of the translation and adherence to the established guidelines across the entire dataset.

\begin{table}[b]
\resizebox{.5\textwidth}{!}{%
\begin{tabular}{|l|l|}
\hline
\rowcolor[HTML]{FFFFC7} 
English  name &
  Arabic name \\ \hline
\multicolumn{1}{|c|}{Wolfgang Reitherman} &
  \begin{tabular}[c]{@{}l@{}} 
        \RL{وولفغانغ رايثيرمان , وولفغانغ ريثيرمان , وولفغانغ ريثرمان}\\
        \RL{ولفغانغ ريتيرمان , وولفغاند رايثيرمان , وولفغانغ ريثيرمن}\\
        \RL{وولفغانغ ريتيرمن , ولفغنغ ريثيرمان , وولفغانغ رايثرمان}\\
        \RL{وولفغنغ رايترمان , وولفغانغ ريتيرمان , ولفغانغ ريثيرمان}\\
  \end{tabular} \\ \hline
\end{tabular}}
\caption{Different ways to write "Wolfgang Reitherman" in Darija}
\label{tab:variants}
\end{table}

\textbf{\dataname} includes four main features: \textbf{db\_id}, representing the database identifier; \textbf{question}, representing the English question; \textbf{darija\_question}, representing the translated question into Moroccan Darija; \textbf{SQL}, the related SQL query; and \textbf{schema}, the database schema, which includes the SQL queries for the creation of all the tables in the related database. An example is displayed in Listing \ref{lst:example}.

\begin{figure}
\begin{lstlisting}[language=SQL, numbers=left, caption={One example of Dialect2SQL}, label={lst:example}]
                `\textbf{\textcolor{blue}{Example}}`
                
schema :
    `\textbf{\textcolor{blue}{CREATE TABLE}}` client (
        client_id   `\textbf{\textcolor{blue}{TEXT}}`   `\textbf{\textcolor{blue}{primary key}}`,
        sex         `\textbf{\textcolor{blue}{TEXT}}`,
        day         `\textbf{\textcolor{blue}{INTEGER}}`,
        address_1   `\textbf{\textcolor{blue}{TEXT}}`,
        address_2   `\textbf{\textcolor{blue}{TEXT}}`,
        district_id `\textbf{\textcolor{blue}{TEXT}}`,
            . . .
        `\textbf{\textcolor{blue}{foreign key}}` (district_id)
            `\textbf{\textcolor{blue}{references}}` district(district_id)
    );
            . . .
            
    `\textbf{\textcolor{blue}{CREATE TABLE}}` events (
        Date received       `\textbf{\textcolor{blue}{DATE}}`,
        Product             `\textbf{\textcolor{blue}{TEXT}}`,
        Timely_response     `\textbf{\textcolor{blue}{TEXT}}`,
        Consumer_disputed   `\textbf{\textcolor{blue}{TEXT}}`,
        Client_ID           `\textbf{\textcolor{blue}{TEXT}}`,
            . . .
        `\textbf{\textcolor{blue}{foreign key}}` (Client_ID)
            `\textbf{\textcolor{blue}{references}}` client(client_id)
    );
            . . .  
            
question :
    What is the full address of the customers who, having received a timely response from the company, have dispute about that response?
        
darija_question : 
    `\scriptsize\<شنو هو العنوان الكامل ديال الكليان اللي، بعد ما وصلهوم>`
    `\scriptsize\< الجواب فالوقت من الشركة، ماعجبهمش داك الجواب؟>`
    
SQL :
    `\textbf{\textcolor{blue}{SELECT}}` 
      T1.address_1, 
      T1.address_2 
    `\textbf{\textcolor{blue}{FROM}}` 
      client `\textbf{\textcolor{blue}{AS}}` T1 
      `\textbf{\textcolor{blue}{INNER JOIN}}` events `\textbf{\textcolor{blue}{AS}}` T2 `\textbf{\textcolor{blue}{ON}}` 
      T1.client_id = T2.Client_ID 
    `\textbf{\textcolor{blue}{WHERE}}` 
      T2.Timely_response = "Yes" 
      `\textbf{\textcolor{blue}{AND}}` T2.Consumer_disputed = "Yes";


\end{lstlisting}
\end{figure}


\subsection{Translation Error}

To illustrate the difference between the automatic and the manual translation, we computed several metrics on automatically translated questions by comparing them to manually translated ones as references. Table \ref{tab:error-rate} presents four main metrics.

\begin{itemize}
  \item \textbf{\textit{CER}} (Character Error Rate), measures the percentage of characters that are incorrect in the translation. Calculated as the number of character insertions, deletions, and substitutions required to convert the translation to the reference, divided by the total number of characters in the reference. \[\text{CER} = \frac{S + D + I}{N}\ = \frac{S + D + I}{S + D + C}\] Where S is the number of substitutions, D is the number of deletions, I is the number of insertions, C is the number of correct characters, N is the number of characters in the reference (N=S+D+C).
  \item \textbf{\textit{WER}} (Word Error Rate), which is similar to CER, but operates in a word level.
  \item \textbf{\textit{TER}} (Translation Edit Rate), measures the number of edits (insertions, deletions, substitutions, and shifts) needed to match the translated text with the reference. It's also normalized by the length of the reference.
  \item \textbf{\textit{CharacTER}} (Character Translation Edit Rate), is a variant of TER that operates at the character level.
\end{itemize}

The results show that, on average, 17\% of the characters in the automatically translated questions are incorrect when compared to the manually translated questions. Also, 23.40\% of the words in the automatically translated questions are inaccurate compared to the manual translations. Finally, the TER score illustrates that 23.30\% is the proportion of changes needed. 

These metrics were computed using HuggingFace library \textbf{\textit{Evaluate}} \footnote{https://huggingface.co/evaluate-metric}

\begin{table}[]
\begin{tabular}{|l|l|l|l|l|}
\hline
\rowcolor[HTML]{FFFFFF} 
\cellcolor[HTML]{FFFFC7}Metric        & \small CER  & \small WER   & \small TER   & \small CharacTER   \\ \hline
\cellcolor[HTML]{FFFFC7}AVG & 0.170 & 0.234 & 0.233  & 0.168 \\ \hline
\end{tabular}
\caption{Average error rates across the translated dataset: Character Error Rate (CER), Word Error Rate (WER), Translation Edit Rate (TER), and Character Translation Edit Rate (CharacTER) }
\label{tab:error-rate}
\end{table}

\subsection{Statistics}

\begin{table*}[]
\centering
\begin{tabular}{|l|l|l|l|l|}
\hline
\rowcolor[HTML]{FFFFC7} 
Database     & N° examples & N° databases & N° examples / db & N° tables / db \\ \hline
BIRD         & 12 751      & 95           & 134              & 7.30            \\ \hline
\dataname & 9 428       & 69           & 137              & 8.00              \\ \hline
\end{tabular}
\caption{\dataname compared to BIRD statistics}
\label{tab:statistics}
\end{table*}

As illustrated in Table \ref{tab:statistics}, \dataname, which is the translated training set of BIRD, consists of 9,428 NLQ-SQL pairs spanning 69 different databases covering diverse domains, such as food, books, education, transport, crime, and more. On average, there are 137 examples per database, though some databases contain only a few dozen examples, while others contain several hundred. Similarly, the number of tables per database varies from 2 to 60, with an average of 8 tables per database. The average number of tables per database in BIRD is 7.30 due to the low complexity of the test set.

\subsection{Baselines}

Large Language Models (LLMs) have rapidely emerged as the best solution for the text-to-SQL task. They have outperformed previous solutions such as rule-based, or sketch-based methods, and traditional machine learning models, by better understanding the questions and their related schemas.

Table \ref{tab:scores} illustrates the performance of three famous families of LLMs dedicated for code generation, StarCoder2 \cite{lozhkov2024starcoder}, Code llama \cite{roziere2023code}, CodeT5 \cite{wang2021codet5}, on a subset of \dataname composed of 697 random questions in the Moroccan dialect.

\begin{table}[]
\begin{tabular}{|l|l|l|l|}
\hline
\rowcolor[HTML]{FFFFC7} 
Model                  & BLEU           & SQAM           & TSED           \\ \hline
\textbf{Starcoder2-7b} & \textbf{0.171} & \textbf{0.403} & \textbf{0.224} \\ \hline
Codellama-7b           & 0.095          & 0.323          & 0.135          \\ \hline
Starcoder2-3b          & 0.086          & 0.335          & 0.031          \\ \hline
CodeT5-2b              & 0.023          & 0.232          & 0.056          \\ \hline
\end{tabular}
\caption{Code based Large Language Models performance on a subset of \dataname}
\label{tab:scores}
\end{table}

In this evaluation, we computed three main metrics, which are defined below.

\begin{itemize} 
    \item \textbf{\textit{BLEU}} (Bilingual Evaluation Understudy), used to evaluate the quality of a generated SQL query compared to one or more reference SQL queries. It compares the n-grams (sequences of n tokens or words) in the generated query to those in the reference queries.
    
    \item \textbf{\textit{SQAM}} (SQL Query Analysis Metric), which divides the predicted and true queries into several clauses (SELECT, FROM, WHERE, etc.) and compares the content of each clause individually, with importance weights assigned to each clause based on its relevance. 
    
    \item \textbf{\textit{TSED}} (Tree Similarity of Editing Distance), a metric that converts both the predicted and true queries into abstract syntax trees (ASTs) and calculates the editing distance between them to capture their structural similarity. 
\end{itemize}

These metrics ranges from 0 to 1, where a higher score indicates higher quality and greater similarity between the queries. 

As shown in Table \ref{tab:scores}, the 7-billion-parameter variant of StarCoder2 outperforms the 7-billion-parameter variant of CodeLlama, as well as the smaller models: the 3-billion-parameter variant of StarCoder2 and the 2-billion-parameter variant of CodeT5. This demonstrates that StarCoder2, particularly in its 7-billion-parameter configuration, offers superior performance in this task compared to both similar-sized and smaller alternatives in the domain of code generation and comprehension.

\vspace{1 cm}
\section{Conclusion \& Future Work}
In this paper, we introduce a novel large-scale, cross-domain text-to-SQL dataset in the Moroccan dialect (Darija), named \dataname. This dataset is manually translated from the English version of BIRD, which is known for its complexity, variety, and the new challenges it introduces in mapping real-world scenarios. To ensure the quality of the dataset, we first perform an initial automatic translation using GPT-4, followed by manual editing of the automatically translated questions by three computer science students who are native speakers of Darija and proficient in SQL. This two-step process, automatic translation followed by detailed manual revision, ensures both linguistic accuracy and alignment with the technical requirements of SQL, thereby enhancing the quality and usability of the dataset. 

While the creation of the first text-to-SQL dataset in an Arabic dialect marks a significant step forward, our journey to improve the performance of text-to-SQL models for Arabic dialects is just beginning. First, we aim to use this dataset to develop a model capable of understanding Darija and performing effectively in the text-to-SQL task. Second, we plan to expand the dataset to include other Arabic dialects, allowing the model to cover a broader range of dialects across the Arabic-speaking world. Finally, we may leverage this dataset to create a translation model capable of translating effectively in both directions, English to Darija and Darija to English, further supporting cross-linguistic applications and bridging the gap between Darija and English-language resources.

\bibliography{references}

\begin{thebibliography}{20}
\providecommand{\natexlab}[1]{#1}

\bibitem[{Almohaimeed et~al.(2024)Almohaimeed, Almohaimeed, Ghanim, and Wang}]{ar_spider_2024}
Saleh Almohaimeed, Saad Almohaimeed, Mansour~Al Ghanim, and Liqiang Wang. 2024.
\newblock \href {https://doi.org/10.1145/3605098.3636065.} {Ar-spider: Text-to-sql in arabic}.
\newblock In \emph{Proceedings of the 39th ACM/SIGAPP Symposium on Applied Computing}, page 1024–1030.
\newblock ArXiv:2402.15012 [cs].

\bibitem[{Bakshandaeva et~al.(2022)Bakshandaeva, Somov, Dmitrieva, Davydova, and Tutubalina}]{pauq_2022}
Daria Bakshandaeva, Oleg Somov, Ekaterina Dmitrieva, Vera Davydova, and Elena Tutubalina. 2022.
\newblock \href {https://doi.org/10.18653/v1/2022.findings-emnlp.175} {Pauq: Text-to-sql in russian}.
\newblock In \emph{Findings of the Association for Computational Linguistics: EMNLP 2022}, page 2355–2376, Abu Dhabi, United Arab Emirates. Association for Computational Linguistics.

\bibitem[{Cao et~al.(2021)Cao, Chen, Chen, Zhao, Zhu, and Yu}]{cao2021lgesql}
Ruisheng Cao, Lu~Chen, Zhi Chen, Yanbin Zhao, Su~Zhu, and Kai Yu. 2021.
\newblock Lgesql: line graph enhanced text-to-sql model with mixed local and non-local relations.
\newblock \emph{arXiv preprint arXiv:2106.01093}.

\bibitem[{Conneau(2019)}]{conneau2019unsupervised}
A~Conneau. 2019.
\newblock Unsupervised cross-lingual representation learning at scale.
\newblock \emph{arXiv preprint arXiv:1911.02116}.

\bibitem[{Dou et~al.(2022)Dou, Gao, Pan, Wang, Che, Zhan, and Lou}]{multi_spider_2022}
Longxu Dou, Yan Gao, Mingyang Pan, Dingzirui Wang, Wanxiang Che, Dechen Zhan, and Jian-Guang Lou. 2022.
\newblock \href {http://arxiv.org/abs/2212.13492} {Multispider: Towards benchmarking multilingual text-to-sql semantic parsing}.
\newblock (arXiv:2212.13492).
\newblock ArXiv:2212.13492 [cs].

\bibitem[{Gao et~al.(2023)Gao, Wang, Li, Sun, Qian, Ding, and Zhou}]{survey_2023}
Dawei Gao, Haibin Wang, Yaliang Li, Xiuyu Sun, Yichen Qian, Bolin Ding, and Jingren Zhou. 2023.
\newblock \href {http://arxiv.org/abs/2308.15363} {Text-to-sql empowered by large language models: A benchmark evaluation}.
\newblock (arXiv:2308.15363).

\bibitem[{Guo et~al.(2021)Guo, Si, Wang, Liu, Fan, Lou, Yang, and Liu}]{chase_2021}
Jiaqi Guo, Ziliang Si, Yu~Wang, Qian Liu, Ming Fan, Jian-Guang Lou, Zijiang Yang, and Ting Liu. 2021.
\newblock \href {https://doi.org/10.18653/v1/2021.acl-long.180} {Chase: A large-scale and pragmatic chinese dataset for cross-database context-dependent text-to-sql}.
\newblock In \emph{Proceedings of the 59th Annual Meeting of the Association for Computational Linguistics and the 11th International Joint Conference on Natural Language Processing (Volume 1: Long Papers)}, page 2316–2331, Online. Association for Computational Linguistics.

\bibitem[{Hazoom et~al.(2021)Hazoom, Malik, and Bogin}]{sede_2021}
Moshe Hazoom, Vibhor Malik, and Ben Bogin. 2021.
\newblock \href {http://arxiv.org/abs/2106.05006} {Text-to-sql in the wild: A naturally-occurring dataset based on stack exchange data}.
\newblock (arXiv:2106.05006).
\newblock ArXiv:2106.05006.

\bibitem[{Huang et~al.(2022)Huang, Wang, Li, Liu, Dou, Yan, Xiao, Wu, and Zhang}]{sesql_2022}
Saihao Huang, Lijie Wang, Zhenghua Li, Zeyang Liu, Chenhui Dou, Fukang Yan, Xinyan Xiao, Hua Wu, and Min Zhang. 2022.
\newblock \href {http://arxiv.org/abs/2208.12711} {Sesql: Yet another large-scale session-level chinese text-to-sql dataset}.
\newblock (arXiv:2208.12711).
\newblock ArXiv:2208.12711 [cs].

\bibitem[{Hui et~al.(2022)Hui, Geng, Wang, Qin, Li, Sun, and Li}]{hui2022s}
Binyuan Hui, Ruiying Geng, Lihan Wang, Bowen Qin, Bowen Li, Jian Sun, and Yongbin Li. 2022.
\newblock S2sql: Injecting syntax to question-schema interaction graph encoder for text-to-sql parsers.
\newblock \emph{arXiv preprint arXiv:2203.06958}.

\bibitem[{Li et~al.(2023)Li, Hui, Qu, Yang, Li, Li, Wang, Qin, Cao, Geng, Huo, Zhou, Ma, Li, Chang, Huang, Cheng, and Li}]{bird_2023}
Jinyang Li, Binyuan Hui, Ge~Qu, Jiaxi Yang, Binhua Li, Bowen Li, Bailin Wang, Bowen Qin, Rongyu Cao, Ruiying Geng, Nan Huo, Xuanhe Zhou, Chenhao Ma, Guoliang Li, Kevin C.~C. Chang, Fei Huang, Reynold Cheng, and Yongbin Li. 2023.
\newblock \href {http://arxiv.org/abs/2305.03111} {Can llm already serve as a database interface? a big bench for large-scale database grounded text-to-sqls}.
\newblock (arXiv:2305.03111).
\newblock ArXiv:2305.03111.

\bibitem[{Lin et~al.(2020)Lin, Socher, and Xiong}]{lin2020bridging}
Xi~Victoria Lin, Richard Socher, and Caiming Xiong. 2020.
\newblock Bridging textual and tabular data for cross-domain text-to-sql semantic parsing.
\newblock \emph{arXiv preprint arXiv:2012.12627}.

\bibitem[{Lozhkov et~al.(2024)Lozhkov, Li, Allal, Cassano, Lamy-Poirier, Tazi, Tang, Pykhtar, Liu, Wei et~al.}]{lozhkov2024starcoder}
Anton Lozhkov, Raymond Li, Loubna~Ben Allal, Federico Cassano, Joel Lamy-Poirier, Nouamane Tazi, Ao~Tang, Dmytro Pykhtar, Jiawei Liu, Yuxiang Wei, et~al. 2024.
\newblock Starcoder 2 and the stack v2: The next generation.
\newblock \emph{arXiv preprint arXiv:2402.19173}.

\bibitem[{Pires(2019)}]{pires2019multilingual}
T~Pires. 2019.
\newblock How multilingual is multilingual bert.
\newblock \emph{arXiv preprint arXiv:1906.01502}.

\bibitem[{Qin et~al.(2022)Qin, Hui, Wang, Yang, Li, Li, Geng, Cao, Sun, Si, Huang, and Li}]{survey_2022}
Bowen Qin, Binyuan Hui, Lihan Wang, Min Yang, Jinyang Li, Binhua Li, Ruiying Geng, Rongyu Cao, Jian Sun, Luo Si, Fei Huang, and Yongbin Li. 2022.
\newblock \href {https://doi.org/10.48550/arXiv.2208.13629} {A survey on text-to-sql parsing: Concepts, methods, and future directions}.
\newblock (arXiv:2208.13629).
\newblock ArXiv:2208.13629 [cs].

\bibitem[{Roziere et~al.(2023)Roziere, Gehring, Gloeckle, Sootla, Gat, Tan, Adi, Liu, Sauvestre, Remez et~al.}]{roziere2023code}
Baptiste Roziere, Jonas Gehring, Fabian Gloeckle, Sten Sootla, Itai Gat, Xiaoqing~Ellen Tan, Yossi Adi, Jingyu Liu, Romain Sauvestre, Tal Remez, et~al. 2023.
\newblock Code llama: Open foundation models for code.
\newblock \emph{arXiv preprint arXiv:2308.12950}.

\bibitem[{Wang et~al.(2019)Wang, Shin, Liu, Polozov, and Richardson}]{wang2019rat}
Bailin Wang, Richard Shin, Xiaodong Liu, Oleksandr Polozov, and Matthew Richardson. 2019.
\newblock Rat-sql: Relation-aware schema encoding and linking for text-to-sql parsers.
\newblock \emph{arXiv preprint arXiv:1911.04942}.

\bibitem[{Wang et~al.(2021)Wang, Wang, Joty, and Hoi}]{wang2021codet5}
Yue Wang, Weishi Wang, Shafiq Joty, and Steven~CH Hoi. 2021.
\newblock Codet5: Identifier-aware unified pre-trained encoder-decoder models for code understanding and generation.
\newblock \emph{arXiv preprint arXiv:2109.00859}.

\bibitem[{Yu et~al.(2019)Yu, Zhang, Yang, Yasunaga, Wang, Li, Ma, Li, Yao, Roman, Zhang, and Radev}]{spider_2019}
Tao Yu, Rui Zhang, Kai Yang, Michihiro Yasunaga, Dongxu Wang, Zifan Li, James Ma, Irene Li, Qingning Yao, Shanelle Roman, Zilin Zhang, and Dragomir Radev. 2019.
\newblock \href {https://doi.org/10.48550/arXiv.1809.08887} {Spider: A large-scale human-labeled dataset for complex and cross-domain semantic parsing and text-to-sql task}.
\newblock (arXiv:1809.08887).
\newblock ArXiv:1809.08887.

\bibitem[{Zhong et~al.(2017)Zhong, Xiong, and Socher}]{wikisql_2017}
Victor Zhong, Caiming Xiong, and Richard Socher. 2017.
\newblock \href {https://doi.org/10.48550/arXiv.1709.00103} {Seq2sql: Generating structured queries from natural language using reinforcement learning}.
\newblock (arXiv:1709.00103).
\newblock ArXiv:1709.00103.

\end{thebibliography}

\end{document}